\documentclass[a4paper]{cas-dc}
\usepackage[numbers]{natbib}
\usepackage{graphicx} 
\usepackage{pgfplots}
\usepackage{enumitem}
\usepackage{amsmath,amssymb,amsfonts}
\usepackage{amsthm}
\newtheorem{theorem}{Theorem}

\theoremstyle{remark}

\usepackage{mathtools}
\usepackage{multirow}
\usepackage{nicefrac}
\usepackage{array}
\usepackage{gensymb}
\usepackage{float}
\usepackage{booktabs} 
\usepackage{stfloats}
\usepackage{leftindex}
\usepackage{easyReview}
\usepackage{hyperref}
\hypersetup{
    colorlinks=true,
    linkcolor=blue,
    citecolor=blue,
    urlcolor=blue
}
\usepackage{cleveref}
 
\pgfplotsset{compat=1.18} 

\begin{document}
\let\WriteBookmarks\relax
\def\textpagefraction{.001}
\shorttitle{Computationally Efficient Volumetric Obstacle Avoidance}
\shortauthors{R. Tapia and I. Soltani}

\title [mode = title] {Relaxed Nonconvex Obstacle Avoidance (RNCOA): Efficient 3D Volumetric Navigation for Autonomous Aerial Vehicles}


\author[1]{Ricardo Tapia}[type=editor,
						auid=000,bioid=1,
						prefix= ,
						role=,
						orcid=0009-0008-0673-2612]
\ead{ricardo.tapia.m@proton.me}

\credit{Conceived the project, designed and carried out the simulations, performed the analysis, and wrote the manuscript}

\author[1]{Iman Soltani}[type=editor,
							auid=000,bioid=1,
							prefix=,
							role=,
							orcid=0000-0001-9430-1522]
\cormark[1]
\ead{isoltani@ucdavis.edu}
\ead[url]{https://lara-soltani.com/}
\credit{Supervision, resources, review and editing}

\affiliation[1]{organization={Department of Mechanical and Aerospace Engineering, University of California-Davis},
	addressline={One Shields Ave}, 
	postcode={95616 CA}, 
	postcodesep={}, 
	city={Davis},
	country={USA}}

\cortext[cor1]{Corresponding author}

\maketitle

\begin{abstract}
The relaxed convex obstacle avoidance (RCOA) formulation is the first approach to generally enable a fully convex optimal control problem (OCP) for obstacle avoidance. Convergence analysis of RCOA yields an analytical framework that defines a unique characteristic: the ability to maintain obstacle avoidance (OA) efficacy even when obstacles reside beyond the controller's prediction horizon. In this paper, RCOA is extended to three-dimensional environments and applied to unmanned aerial vehicle (UAV) navigation. Furthermore, the formulation is enhanced to account for the area/volume of the vehicle, moving beyond point-geometry representations to enable collision avoidance between 3D objects. Numerical simulations demonstrate that RCOA yields computational performance on par with or exceeding state-of-the-art methods. Notably, RCOA is shown to enable a nonlinear model predictive controller (NMPC) to execute aggressive maneuvers through narrow passages with reduced prediction horizons, ensuring real-time feasibility at frequencies exceeding 30~Hz. Selected simulations are presented at \href{https://rtapia8484.github.io/RCOA/}{rtapia8484.github.io/RCOA}.
\end{abstract}

\section{Introduction}
The proliferation of UAVs in highly constrained, unstructured environments, ranging from urban air mobility and autonomous delivery to search and rescue operations, has intensified the demand for robust local path planning. Autonomy in these scenarios requires not only navigating unforeseen environmental changes and dynamic obstacles, but doing so while pushing the vehicle to its dynamic limits. Local planners must generate strictly collision-free trajectories that respect high-order nonlinear vehicle dynamics and satisfy external constraints, such as actuator saturation or wind disturbances \cite{Romero2022}.

Existing methodologies are broadly categorized into non-optimization and optimization-based approaches. Classic non-optimization methods, such as artificial potential fields (APF) \cite{SuryaPrakash2025}, hybrid A*, and the dynamic window approach (DWA) \cite{Dobrevski2024}, are computationally efficient but often suffer from local minima or a lack of dynamic consistency. Recent efforts attempt to mitigate these issues by integrating sampling-based methods like RRT with DWA or Reinforcement Learning \cite{Han2025, Allen2019}. However, multi-component frameworks often increase architectural complexity without addressing the underlying limitations of the individual algorithms. Sampling-based methods particularly struggle with high-dimensional differential constraints, often requiring the solution of complex Boundary Value Problems (BVP) or excessive linearization to maintain internode accuracy \cite{Perez2012}.

Optimization-based methods, specifically model predictive control (MPC) and NMPC, inherently integrate dynamics and constraints. Although real-time applications are often hindered by the nonconvexity of vehicle dynamics and the computational burden of OA constraints, increasing computational power makes NMPC more tractable. A significant limitation of standard NMPC with OA constraints is the dependence on the prediction horizon. If an obstacle is outside this horizon, the controller remains "blind," necessitating longer horizons that increase computational cost and may violate real-time frequency requirements.

This creates a severe operational paradox for high-speed, agile systems. To navigate in cluttered environments safely, a fast-moving UAV requires a prediction horizon long enough to detect obstacles and compute feasible evasive maneuvers. However, extending this horizon exponentially increases computational burden, particularly when evaluating complex, nonconvex OA constraints. If computational latency exceeds the sampling time, the controller destabilizes. Conversely, keeping the horizon short to maintain high frequencies renders the NMPC effectively myopic. A myopic controller operating with hard spatial constraints will not react until an obstacle enters its horizon, often resulting in inevitable collisions due to actuator limits and inertia. Breaking this dependency between prediction horizon and avoidance capability is critical for deploying real-time NMPC in agile robotics.

Furthermore, several OA formulations simplify the representation of the vehicle to point-geometry, and the physical dimensions of the vehicle are typically accounted for by uniformly inflating the obstacle boundaries. While mathematically convenient, static inflation is highly conservative and orientation-independent. For a dynamically rotating UAV, approximating the vehicle as a bounding sphere eliminates the ability to execute aggressive, high-roll maneuvers to slip through narrow gaps. High-performance autonomy requires the OCP to understand the exact, rotating volumetric footprint of the vehicle to leverage attitude and thread through tight corridors that point-mass formulations would classify as infeasible.

To address these challenges, the relaxed convex obstacle avoidance (RCOA) formulation was presented in \cite{Tapia2025}. Initially validated in 2D using bicycle models with nonlinear tire dynamics, RCOA outperformed standard formulations like \cite{Chen2017} and \cite{Schouwenaars2001}. This paper extends RCOA as follows:
\begin{enumerate}
	\item \textbf{3D Scaling:} Extends the RCOA formulation to three-dimensional space to address scaling challenges inherent in aerial navigation.
	\item \textbf{Vehicle Geometry:} Transitions from point-geometry representations to a multi-point approximation, allowing high-fidelity collision checking between 3D objects without excessive conservatism. 
	\item \textbf{Horizon Independence:} Demonstrates RCOA's unique analytical property, which provides a guidance effect for obstacles residing outside the NMPC prediction horizon in three-dimensions.
\end{enumerate}

These new OA additions are evaluated through two primary quadrotor simulations. The first assesses 3D navigation efficiency with varying vehicle representations (area/volume) against notable OA benchmarks, demonstrating RCOA's consistent computational advantage. The second highlights a high-performance maneuver, navigating through a narrow passage under aggressive roll angles. The latter simulation shows that RCOA facilitates successful navigation with a reduced prediction horizon, enabling controller frequencies over 30~Hz despite nonconvex underlying dynamics.

The remainder of this paper is organized as follows: Section \ref{sec: Related Work} reviews related work on optimization-based OA. Section \ref{sec: Methodology} defines the 3D RCOA formulation and the representation of the vehicle to multi-point to approximate area/volume. Section \ref{sec: simulations} presents the simulation results, and Section \ref{sec: Conclusion} concludes the paper.

\section{Related Work} \label{sec: Related Work}
    \subsection{Obstacle Avoidance}

The obstacle avoidance problem is foundational across robotics, with methodologies broadly categorized into non-optimization and optimization-based approaches. Early non-optimization methods, such as artificial potential fields (APF) \cite{Khatib1985, Borenstein1990} and the Gilbert-Johnson-Keerthi (GJK) algorithm \cite{Gilbert1988}, paved the way for geometric collision checking.  The subsequent integration of these concepts into optimal control has driven significant progress. For instance, APF methods \cite{She2025, LiuXiaofengandChen2018} and ellipsoidal calculus \cite{Ros2002, Rosenfelder2025, Leprich2025} have been directly incorporated into an OCP. 
     
Naturally, most OA formulations are based on defining spatial regions of the obstacle, typically simplified to primitive shapes such as rectangles and ellipses in \(\mathbb{R}^2\) \cite{Schouwenaars2001,Chen2017}, or polytopes and ellipsoids in \(\mathbb{R}^3\) \cite{Tracy2023,Pereira2021}. Some combine both Euclidean distance and spatial regions \cite{Pereira2021,Zhang2021}.  The alternative approach is to define obstacle-free regions instead \cite{Morozov2024, Garg2025}. The focus here is on OA formulations that use primitive shapes to define obstacles, specifically, formulations that naturally extend to three dimensions and can also account for the shape (volume) of the controlled vehicle. 

Recent research has yielded significant advances in OA formulations that capture vehicle volume. A notable approach is DCOL \cite{Tracy2023}, which extends the foundational work of \cite{Gilbert1994} by embedding the minimum scaled distance optimization problem directly in the OCP. DCOL accommodates a variety of primitive shapes, from polytopes to ellipsoids. While it exhibits strong performance, it inherently requires solving a convex minimum scaled distance problem at every iteration and passing differentiable KKT conditions to the OCP solver. Consequently, achieving real-time tractability with this framework heavily relies on highly optimized, language-specific custom cone solvers, as standard solvers generally yield significantly higher computational latency for this specific architecture. 

Another prominent formulation is CIOA \cite{SCHOELS20206555}; the formulation has a convex algorithm but essentially requires the calculation of the obstacle-free space a priori using a series of circular bounds. This is achieved by determining an initial obstacle-free path, which can be done using RRT. This algorithm would not pose a problem in linear cases, but if nonlinear dynamics are present in the OCP, then the linear solutions of RRT and OCP may result in large deviations. Large deviations will require a substantial number of small optimization problems to be solved a priori, likely necessitating a lookup table.

Other prominent volumetric formulations include the application of ellipsoidal calculus \cite{Rosenfelder2025} and dual problem approaches capable of accommodating polytopic or spherical vehicle representations \cite{Zhang2021}. To evaluate the overlap of two ellipsoids, \cite{Rosenfelder2025} define a scaled combination of two ellipsoids, leveraging the criteria established in \cite{Gilitschenski2014}:
\begin{equation} \label{eq: Ellipsoidal K}
    K(\lambda)=1-(w-v)^T\bigg(\frac{1}{(1-\lambda)}\textbf{B}^{-1}+\frac{1}{\lambda}\textbf{A}^{-1}\bigg)^{-1}(w-v)
\end{equation}
where $\mathbf{A}$ and $\mathbf{B}$ are positive definite matrices defining the geometries of the respective ellipsoids, $v, w \in \mathbb{R}^n$ denote their centers, and $\lambda \in [0,1]$ is a scaling factor. Collision avoidance is guaranteed if there exists a $\lambda \in [0,1]$ such that $K(\lambda)\leqslant 0$. When applied within an OCP, the center position of the vehicle $v$ and the rotation matrix (embedded in $\mathbf{A}$) become decision variables, the formulation becomes highly nonconvex. Notably, it's fundamentally similar to the standard point-to-ellipsoid OA constraint of \cite{Chen2017}:
\begin{equation} \label{eq: EOA def}
    1-(v-w)^T \mathbf{B}(v-w) \leqslant\:0 
\end{equation}

Alternatively, the formulation proposed in \cite{Zhang2021} utilizes the dual problem of the minimum distance between two convex sets \cite{Boyd2014ConvexOptimization}. Defining the obstacle set as $\mathbf{A}_o y_o \leqslant b_o$ and the vehicle set as $\mathbf{A}_v\mathbf{R}_b^i(y_v-v) \leqslant b_v$ (where $\mathbf{R}_b^i \in SO(3)$ maps the body frame to the inertial frame), the resulting dual OA constraints are given by:
\begin{equation} \label{eq: DMDOA def}
    \begin{aligned}
        -b_v^T\mu+(\mathbf{A}_o v -b_o)^T\lambda & > 0 \\
        -\mathbf{A}_v^T \mu + \mathbf{R}^T \mathbf{A}_o^T \lambda & = 0 \\
        \lVert \mathbf{A}_o^T \lambda\rVert_* = 1,\quad \lambda\geqslant 0,\quad \mu &\geqslant 0 
    \end{aligned}
\end{equation}
where $\mu$ and $\lambda$ are dual variables. Enforcing the dual norm constraint $\lVert \mathbf{A}_o^T \lambda\rVert_* = 1$ ensures that the spatial distance between the two sets is actively minimized.

This paper evaluates RCOA against the formulations defined in \eqref{eq: Ellipsoidal K}-\eqref{eq: DMDOA def}. Specifically, the RCOA multi-point extension is compared with two 3D volumetric formulations. To analyze the performance of the multi-point representation at both a micro and macro level, this same multi-point extension is also applied to \eqref{eq: EOA def}. For \eqref{eq: Ellipsoidal K}, the continuous $\lambda \in [0,1]$ representation from \cite{Rosenfelder2025} is chosen over the discrete optimization of $\lambda^*$ \cite{Gilitschenski2014} to increase the OCP solver's degrees of freedom. Notably, none of these methods require substantial pre-processing to generate obstacle-free trajectories.
    
\section{Methodology} \label{sec: Methodology}

\subsection{RCOA} \label{subsec: obstacle avoidance}

The foundational RCOA formulation \cite{Tapia2025} defines an obstacle boundary as a rectangular region defined by two vertices, $(x^o_{\min}, y^o_{\min})$ and $(x^o_{\max}, y^o_{\max})$. The core logic of the OA formulation is defined by the following spatial exclusions:
\begin{equation}\label{RCOA logic def 1}
	x^o_{\min} \leqslant X \leqslant x^o_{\max} \implies Y \geqslant y^o_{\max}
\end{equation}
\begin{center}
	\textit{or}
\end{center}
\begin{equation}\label{RCOA logic def 2}
	x^o_{\min} \leqslant X \leqslant x^o_{\max} \implies Y \leqslant y^o_{\min} 
\end{equation}
where $(X,Y)$ designates the Cartesian position of the vehicle. Assuming that the obstacle is aligned with the $(X,Y)$ axes, a vehicle bounded laterally by the obstacle must remain strictly above or below it. These logic conditions can be mathematically represented using the big-M method \cite{Bemporad1999ControlConstraints} with integer variables $\gamma_I \in \mathbb{Z}$.

The resulting OA mixed integer linear programming (MILP) constraint(s) are: 
\begin{subequations} \label{eq: RCOA 2D MIP}
	\begin{align}
		-X &\leq -x^o_{\text{min}} + M_1 \gamma_{I_1} \label{convexCons1a} \\
		X &\leq x^o_{\text{max}} + M_2 \gamma_{I_2} \label{convexCons1b} \\
		Y &\geq y^o_{\text{max}} - M_3(\gamma_{I_1} + \gamma_{I_2}) \label{eq: RCOA MIP Ymax} \\
		\textbf{or} \quad Y & \leqslant y^o_{\min} + M_3 (\gamma_{I_1} + \gamma_{I_2}) \label{eq: RCOA MIP Ymin} \\
		\gamma_{I_1} + \gamma_{I_2} &\leq 1,\quad \gamma_{I_i} \in \{0,1\} \label{convexCons1d}
	\end{align}
\end{subequations}
\noindent where the spacial homotopy is divided into two, above $Y_{\max}$ \eqref{eq: RCOA MIP Ymax} \textbf{or} below $Y_{\min}$ \eqref{eq: RCOA MIP Ymin}. Incorporating these constraints into to an MILP OCP, results in two possible independent problems, defined by applying  \eqref{eq: RCOA MIP Ymax} \textbf{or} \eqref{eq: RCOA MIP Ymin}. Each problem can generate a feasible trajectory; thus, the globally optimal solution is given by the problem that generates the lowest cost. 
	
Relaxing the integral constraints from discrete to continuous, and adding a linear penalty function for the relaxed variables results in the following OCP,
\begin{subequations}  \label{eq: RCOA 2D}
	\begin{align}
		\min \quad &f_0 + w(\gamma_1 + \gamma_2)\\ 
		\text{s.t.}\;  c_E &= 0\\
		\quad -X &\leqslant -x^o_{\min} + M_1 \gamma_1    \\
		 X &\leqslant x^o_{\max} + M_2 \gamma_2   \\ 
		Y & \geqslant y^o_{\max} - M_3 (\gamma_1 + \gamma_2) \label{eq: RCOA 2D ymax} \\
		\textbf{or} \quad Y & \leqslant y^o_{\min} + M_3 (\gamma_1 + \gamma_2) \label{eq: RCOA 2D ymin} \\
		\gamma_1 + \gamma_2 & \leqslant 1, \quad \gamma_1, \gamma_2 \in [0,1] \\
		c_I &\leqslant 0
	\end{align}
\end{subequations}
\noindent where $f_0$ is the primary cost function, $c_E$ and $c_I$ are equality and inequality constraints respectively, the penalty weight $w>0$, $M_i \in \mathbb{R}_{++}$ are sufficiently large constants, and $\gamma_i$ serve as relaxed continuous variants of binary indicator variables. The OCP is convex by \cite[Remark 1]{Tapia2025} if the cost function of $f_0$ is convex, and the constraints of $c_E$ and $c_I$ are convex. 

Assuming feasibility, the OA formulation functions as follows: If the vehicle approaches the left obstacle boundary ($x_{\min}^o$), the penalty function drives $\gamma_1 \to 0$, effectively removing the $M_1\gamma_1$ buffer. Once the conditional domain $x^o_{\min} \leqslant X \leqslant x^o_{\max}$ is entered, the penalty function drives $\gamma_1 = \gamma_2 = 0$. Consequently, when $\gamma_1=0\text{ and }\gamma_2 = 0$, they enforce $Y \geqslant y^o_{\max}$ via \eqref{eq: RCOA 2D ymax} or $Y \leqslant y^o_{\min}$ via \eqref{eq: RCOA 2D ymin}. As with \eqref{eq: RCOA 2D MIP}, solving for the globally optimal path requires formulating two OCPs that are independent.

A defining characteristic of RCOA emerges from its optimality conditions. As derived from KKT analysis in \cite[Theorem 1]{Tapia2025}, the optimal spatial states satisfy the following relationship:
\begin{equation} \label{eq: RCOA OPT RESULT}
    Y^*\geqslant y^o_{\max}-\frac{M_3}{M_1}(x^o_{\min}-X^*)
\end{equation}
This analytical property dictates that as the vehicle approaches the lateral boundary ($x^o_{\min}$), its vertical position is proportionally regulated toward the safe boundary ($y^o_{\max}$). Crucially, this KKT derived regulation acts independently of the OCP prediction horizon. It provides an intrinsic anticipatory guidance, prompting evasive action before the obstacle enters the finite horizon window, a unique advantage over hard-constrained formulations that remain blind beyond the prediction horizon.

\subsection{Three-Dimensional and Volumetric RCOA Extensions}
To extend RCOA to three-dimensional applications, obstacles are modeled as rectangular prisms and defined by a local coordinate frame. The obstacle set is defined as $\mathcal{O}=\{x\in \mathbb{R}^3 \mid \mathbf{A}x\leqslant b\}$, where $\mathbf{A} \in \mathbb{R}^{6\times 3}$ contains orthogonal normal vectors corresponding to the fixed obstacle coordinate system ($\pm \hat{x}, \pm \hat{y}, \pm \hat{z}$), and $b$ defines the physical dimensions, i.e. half lengths of, $b=(x^o_{\max}, -x^o_{\min}, y^o_{\max}, -y^o_{\min}, z^o_{\max}, -z^o_{\min})^T$. The vehicle's position relative to the obstacle is obtained via the homogeneous transformation matrix in $SE(3)$:
\begin{equation} \label{eq: Transformation matrix}
    T = \begin{bmatrix}
        \mathbf{R}_{OW} & p \\
        0 & 1
    \end{bmatrix}
\end{equation} 
where $\mathbf{R}_{OW} \in SO(3)$ maps the world frame to the obstacle frame, and $p$ translates the world origin.

Because RCOA fundamentally relies on domain partitioning, it naturally scales to three dimensions. Consider the Euclidean space of $ \boldsymbol{S} \in \mathbb{R}^3$ and $\boldsymbol{Q}\in \mathbb{R}^2$, with $\boldsymbol{Q}$ being a subset of $\boldsymbol{S}$, i.e. $\boldsymbol{Q} \subset \boldsymbol{S}$. Letting the vehicle position in the world frame be defined by $\boldsymbol{x} \in \mathbb{R}^3$, then the vehicle position relative to the obstacle frame is defined by $\tilde{\boldsymbol{x}} \in \mathbb{R}^3$, calculated as $[\tilde{\boldsymbol{x}},1]^T = T \cdot [\boldsymbol{x},1]^T$. Here, the space $\boldsymbol{S}$ is defined by the obstacle coordinate system, and any two orthogonal axes are isolated as the subspace of $\boldsymbol{Q}$ and define RCOA in this subspace. 

For example, consider the obstacle illustrated in Figure~\ref{fig: RCOA Obstacle Def}. The subspace of $\boldsymbol{Q}$ is defined by any two coordinates, say $(x_1,y_1)$, applying RCOA with $(X=x_1,Y=y_1)$ will result in the OCP \eqref{eq: RCOA 2D} after the necessary transformations. A possible trajectory is also shown in the figure, where the green planes define the desired trajectory space (capped by the obstacle surface normal to $+y_1$) when $\tilde{\boldsymbol{x}}\in[x_{\min}^o,x_{\max}^o]$. The planes are perpendicular to $x_1$ and positioned at $x_{\min}$ and $x_{\max}$. Note that the feasible set is unconstrained in the $z_1$-axis, extending along its entire 1D-subspace. Since the obstacle is defined as a prism, it's convenient to define an axis-aligned coordinate system, i.e. axis are normal to the intersecting surface. 

To complete RCOA's extension to three-dimensions, Theorem 1 from \cite{Tapia2025} is restated and extended into the full Euclidean space of $\boldsymbol{S}$ below. 

\begin{figure}[t] 
	\centerline{\input{RNCOA_Lateral.pgf}}
	\vspace{-0.5cm}
	\caption{Typical obstacle with coordinate frame centered and axis normal to surface. The trajectory in green corresponds to an RCOA OCP applied to the plane formed by $(x_1,y_1)$.}
	\label{fig: RCOA Obstacle Def}
\end{figure}

\begin{theorem}[Conditional Statement Equivalence] \label{Theorem 1}
	Let a centered axis-aligned obstacle coordinate system be defined by $\mathcal{C}=\{x,y,z\}$, then RCOA is mapped onto the subspace $Q\in\{x_1,x_2\}$ defined by any two orthogonal axes $(x_1,x_2)$, with $X=x_1$ and $Y=x_2$, where $x_1,x_2\in \mathcal{C}$ and $x_1\neq x_2$. Under assumptions A1–A4, the continuous RCOA formulation contained in \eqref{eq: RCOA 2D} exhibits conditional equivalence to the mixed-integer constraints of \eqref{eq: RCOA 2D MIP}. Specifically, the continuous relaxation strictly enforces identical spatial avoidance boundaries (e.g., $Y \geqslant x^o_{2,\max}$ when $X \in [x^o_{1,\min}, x ^o_{1,\max}]$) as the original discrete logical conditions at the optimal solution.
\end{theorem}

\begin{enumerate}[label=\textup{(A\arabic*)},ref=\textup{A\arabic*}]
	\item An obstacle-free trajectory is feasible.\label{Asumption A: 1}
	\item The cost function of $f_0$ is convex and penalizes path deviation with unit weight. \label{Asumption A: 2}
	\item The system dynamics do not depend explicitly on $(X,Y)$, applicable to common automobile and quadrotor $(X,Y,Z)$ models. Additionally, the system dynamics are affine.
	\item $M_i$ are chosen sufficiently large to preserve feasibility when $\gamma_i=1$ and the upper bounds on $\gamma_i$ are inactive. \label{Asumption A: 4}
\end{enumerate} 

The assumptions ensure feasibility and simplify the analysis. Consider a vehicle traveling along the $+X$-axis, i.e., the reference path, approaching a rectangular obstacle centered at the origin. The OA trajectory is obtained from the OCP of \eqref{eq:P1}, where $y_n\in \mathbb{R}^n$ includes position variables $(X,Y,Z)$, $c_{E}$ and $c_{I,*}$ are equality and inequality constraints respectively, with $*\in \{1,2,\dots,8\}$. Assumption \ref{Asumption A: 2} and Remark~\cite[Remark 1]{Tapia2025} guarantee the OCP is convex. 

\begin{equation} \tag{P1} \label{eq:P1}
	\begin{aligned}
		\min\limits_{y_n,\gamma} \,&f_0(y_n)\:+\: w(\gamma_1 + \gamma_2) \\
		s.t.\quad&c_E(y_n) = 0 \\
		\text{c}_{I,1}:\quad& -X \leqslant -x^o_{1,\text{min}} + M_1 \gamma_1  \\ 
		\text{c}_{I,2}:\quad& X \leqslant x^o_{1,\text{max}} + M_2 \gamma_2  \\
		\text{c}_{I,3}:\quad& Y \geqslant x^o_{2,\text{max}} - M_3 (\gamma_1 + \gamma_2)  \\
		\text{c}_{I,4/5}:\quad& -\gamma_1 \leqslant 0,\;-\gamma_2 \leqslant 0 \\
		\text{c}_{I,6/7}:\quad& \gamma_1 \leqslant 1,\;\gamma_2 \leqslant 1 \\
		\text{c}_{I,8}: \quad& \gamma_1 + \gamma_2-1\leqslant0 
	\end{aligned}
\end{equation}

\renewcommand{\proof}{\textbf{Proof. }}
\noindent \begin{proof}
	The proof is provided in \cite{Tapia2025} and applies equivalently to any subspace $\boldsymbol{Q}$ by letting $X=x_1$ and $Y=x_2$. 
    
    For complex geometry such as corners, with the constraint of \eqref{eq: RCOA corner setup} added to \eqref{eq:P1}, RCOA is extended to the full Euclidean space $\boldsymbol{S}$,
	\begin{subequations} \label{eq: RCOA corner setup}
		\begin{align}
			Z & \leqslant x^o_{3,\text{min}} + M_4 (\gamma_1 + \gamma_2)\label{eq: RCOA Z max} \\
			\textbf{or}\; Z & \geqslant x^o_{3,\text{max}} - M_4 (\gamma_1 + \gamma_2). \label{eq: RCOA Z min} 
		\end{align}
	\end{subequations} 
	The $Z$-axis constraint can be applied without introducing new variables. Assuming the following constraint is added to the Lagrangian of \eqref{eq:P1}, $c_{I,9}: Z\geqslant x_{3,\max}^o-M_4(\gamma_1+\gamma_2)$, with $\lambda_9$ as the corresponding dual variable. The Karush-Khun-Tucker (KKT) stationary conditions \cite{Boyd2014ConvexOptimization} with respect to $Y$, $Z$, and $\gamma_i$ are,
	\begin{subequations} \label{eq: primal lagrange KKT}
		\begin{align}
			\mathcal{L}_Y  &= \nicefrac{\partial f_0 }{\partial Y}-\lambda_3  =0  \label{eq: primal lagrange KKT, Y} \\
			\mathcal{L}_Z  &= \nicefrac{\partial f_0 }{\partial Z}-\lambda_9  =0  \label{eq: primal lagrange KKT, Z} \\
			\mathcal{L}_{\gamma_1} &= w-\lambda_1 M_1-\lambda_3 M_3 -\lambda_4 + \lambda_8-\lambda_9 M_4 = 0 \label{eq: primal lagrange KKT, gamma1} \\
			\mathcal{L}_{\gamma_2} &= w-\lambda_2M_2-\lambda_3 M_3 - \lambda_5 + \lambda_8 -\lambda_9 M_4=0 \label{eq: primal lagrange KKT, gamma2} 
		\end{align}
	\end{subequations}
	where $Z=x_3,\,x_3\in \mathcal{C}$, $x_3\neq x_1\text{, and } x_3\neq x_2$. Since the vehicle is traveling along the $+X$-axis, the analysis proceeds with the same simplifications as in \cite{Tapia2025}, i.e. $\lambda_4^*=0,\;\lambda_8^*=0$, $\gamma_2^*\in[0,\epsilon]$. This results in the following stationary conditions for $\gamma_1,\,\gamma_2$:
	\begin{subequations} \label{eq: primal lagrange KKT gamma}
		\begin{align}
		\lambda_1^* M_1+\lambda_3^* M_3 + \lambda_9^* M_4 &= w \label{eq: primal lagrange KKT gamma1}\\
		\lambda_3^* M_3 + \lambda_5^* + \lambda_9^* M_4 &= w \label{eq: primal lagrange KKT gamma2}
		\end{align}
	\end{subequations} 
	which imply $\lambda_1^*M_1=\lambda_5^*$. In terms of $\gamma_1$, Analyzing when the constraints of $c_{I,1},\,c_{I,3}$, and $c_{I,9}$ are active, yields the following cases:
	
	\textbf{Case 1 (($\gamma_1$ from $\text{c}_{I,3}$) or ($\gamma_1$ from $\text{c}_{I,9}$) is larger, $\lambda_1^*=0$):} If $c_{I,3}$ is active, then due to complementary slackness $\lambda_3^*>0,\,\lambda_9^*=0$, and from \eqref{eq: primal lagrange KKT, Y}, \eqref{eq: primal lagrange KKT, Z}, and \eqref{eq: primal lagrange KKT gamma} yield:
	\begin{equation*} \label{proof: case 1 opt 1}
		\partial f_0/\partial Y=\lambda_3^*,\;\partial f_0/\partial Z=0,\; \lambda_3^* M_3=w, \; \lambda_5^*=0
	\end{equation*}
	else if $c_{I,9}$ is active :
	\begin{equation*} \label{proof: case 1 opt 2}
		\partial f_0/\partial Y=0,\;\partial f_0/\partial Z=\lambda_9^*,\; \lambda_9^* M_4=w, \quad \lambda_5^*=0
	\end{equation*}
	with $\lambda_9^*>0,\,\lambda_3^*=0$, and $\lambda_5^*=0$ implying $\gamma_2^*=\epsilon$.
	
	\textbf{Case 2 (The expressions are equal):} $c_{I,1},\,c_{I,3},$ and $c_{I,9}$ are active and imply that $(\lambda_1^*,\lambda_3^*,\lambda_9^*)> 0$. Substituting \eqref{eq: primal lagrange KKT, Y} and \eqref{eq: primal lagrange KKT, Z} into \eqref{eq: primal lagrange KKT gamma1} yields:
	\begin{align*}
		\lambda_1^* M_1 + \nicefrac{\partial f_0 }{\partial Y}M_3 + \nicefrac{\partial f_0 }{\partial Z}M_4=w
	\end{align*}
    which implies $w>\nicefrac{\partial f_0 }{\partial Y}M_3 + \nicefrac{\partial f_0 }{\partial Z}M_4$ to enforce obstacle avoidance, and $\lambda_5^*>0\rightarrow \gamma_2^*=0$. Substituting $c_{I,1}$ into both $c_{I,3}$ and $c_{I,9}$ yield the following expressions:
    \begin{align*}
    		Y^* &= x_{2,\max}^o-\nicefrac{M_3}{M_1}(x^o_{1,\min}-X^*) \\
    		Z^* &= x_{3,\max}^o-\nicefrac{M_4}{M_1}(x^o_{1,\min}-X^*) 
    \end{align*}
    
    \textbf{Case 3 ($\gamma_1$ from $\text{c}_{I,1}$ is larger):} Dictates $\lambda_1^*>0$ with $(\lambda_3^*,\lambda_9^*)=0$, evaluating \eqref{eq: primal lagrange KKT, Y}, \eqref{eq: primal lagrange KKT, Z}, and \eqref{eq: primal lagrange KKT gamma} yield:
    \begin{equation*} \label{proof: case 3 opt}
    	\partial f_0/\partial Y=0,\;\partial f_0/\partial Z=0,\; \lambda_1^* M_1=w, \; \lambda_5^*=w
    \end{equation*}
   	Concurrently, substituting $c_{I,1}$ into $c_{I,3}$ and $c_{I,9}$ for $\gamma_1$ yields:
    \begin{subequations} \label{eq: case 3 opt YZ}
    	\begin{align}
    		Y^* &> x_{2,\max}^o-\nicefrac{M_3}{M_1}(x^o_{1,\min}-X^*) \label{eq: case 3 opt Y}\\
    		Z^* &> x_{3,\max}^o-\nicefrac{M_4}{M_1}(x^o_{1,\min}-X^*) \label{eq: case 3 opt Z}
    	\end{align}
    \end{subequations}
    
    These optimal conditions mirror the two-dimensional case. Case 1 characterizes the need to deviate from the reference path in either $Y$ or $Z$. Case 2 is likely to occur as a transition to case 3, as it requires three inequalities to be simultaneously active. Case 3 minimizes path deviation while ensuring obstacle avoidance. This means the primal cost is minimized and the dual function is maximized ($M_i>1$). 
    
    Thus, when approaching the obstacle from $+X$-axis, the optimal solution of \eqref{eq:P1} will generally follow \eqref{eq: case 3 opt YZ}, ensuring $Y\geqslant x_{2,\max}^o$ and $Z\geqslant x_{3,\max}^o$ when $X \in [x_{1,\min}^o,x_{1,\max}^o]$. By formulation symmetry, the same analysis applies when approaching the obstacle from $-X$-axis or when applying \eqref{eq: RCOA 2D ymin} or \eqref{eq: RCOA Z min}, completing the proof of conditional equivalence.   
    
\end{proof}

Note the linear relationship of the lateral position ($Y$,$Z$) to the relative distance between the vehicle and obstacle still applies in $\boldsymbol{S}$, and is independent of the prediction horizon. 

\subsubsection{Volumetric Approximation} \label{subsubsection}

The formulation as presented assumes a point-geometry configuration for the controlled vehicle, requiring conservative inflation of the obstacle to account for vehicle volume. High-performance applications require high-fidelity geometry. To address this, the vehicle's volume is approximated by a set of vertices, as illustrated in Figure \ref{fig: Volumetric Representation}. This includes two- and three-dimensional configurations, which can be rigidly fixed to the world/vehicle frame or dynamically rotate in the body frame.

\begin{figure}[ht]  
    \centerline{\input{VolumetricRep.pgf}}
    \vspace{-.5cm}
    \caption{Area/volumetric representation of the vehicle (quadrotor): A. Vehicle fixed area defined by four vertices, B. Body fixed rotating area defined by four vertices, C. Vehicle fixed cube defined by eight vertices, D. Body fixed rotating prism defined by eight vertices.}
    \label{fig: Volumetric Representation}
\end{figure}

Applying the standard RCOA constraints identically to all vertices of representation $D$ (Figure \ref{fig: Volumetric Representation}) introduces a severe computational penalty, dramatically increasing the number of constraints and variables. While acceptable in strictly convex settings, this increase in complexity is detrimental in nonconvex programming. To maintain computational tractability while ensuring rigorous collision avoidance, the following nonconvex multi-point formulation is presented:
\begin{subequations} \label{eq: RNCOA def}
	\begin{align}
		-\max_{i \in \mathcal{V}} \{X_i\} &\leqslant -x^o_{1,\min} + M_1 \gamma_1 \label{eq: RCOA nonc xmin} \\ 
		\min_{i \in \mathcal{V}} \{X_i\} &\leqslant x^o_{1,\max} + M_2 \gamma_2   
		\label{eq: RCOA nonc xmax} \\ 
		Y_i &\geqslant x^o_{2,\max} - M_3 (\gamma_1 + \gamma_2) \quad \forall i \in \mathcal{V} \label{eq: RCOA nonc ymax} \\ 
		\text{or } Y_i &\leqslant x^o_{2,\min} + M_3 (\gamma_1 + \gamma_2) \quad \forall i \in \mathcal{V} \label{eq: RCOA nonc zmax} \\
		\gamma_1 + \gamma_2 &\leqslant 1,\quad \gamma_1, \gamma_2 \in [0,1] \label{eq: RCOA nonc gamma} 
	\end{align}
\end{subequations}
where $\mathcal{V}$ is the set of all vehicle vertices $i=1,\dots,N_{v}$, the obstacle coordinate system is defined by $\mathcal{C}=\{x,y,z\}$, $(x_1,x_2)\in\mathcal{C}$ with $X=x_1$, $Y=x_2$, and $x_1\neq x_2$. Although determining the bounding extents ($\max\{X_i\}$ and $\min\{X_i\}$) sacrifices strict convexity, it dramatically reduces the constraint matrix size and avoids additional variables. These non-differentiable functions can be handled via the subgradient method \cite{Shor1985}.

This modification can be easily incorporated into the KKT optimal conditions in Theorem~\ref{Theorem 1}. Furthermore, the underlying conditional logic conservatively requires only a single vertex to cross the $[x_{1,\min}^o,x_{1,\max}^o]$ threshold to trigger the $x_{2}^o$ clearance constraints for the entire vehicle body. The constraint set of \eqref{eq: RNCOA def} along with the linear penalty function of $w(\sum \gamma_i)$ is termed the relaxed nonconvex obstacle avoidance (RNCOA) formulation to differentiate from the convex version. 

Because navigating an obstacle in 3D offers multiple valid spatial corridors (e.g., over, under, left, right), R(N)COA naturally fits within a hierarchical planning architecture. A global planner can dictate the preferred spatial corridor, example RRT, while the local RCOA-NMPC selectively solves the corresponding OCP to generate the dynamically optimal evasion maneuver.  Note, as stated in \cite{Tapia2025} to solve for the globally optimal one would solve every possible configuration in parallel to determine the minimum cost, but any one configuration can generate a valid obstacle-free path. 

The simulations presented in Section \ref{sec: simulations} evaluate this volumetric RNCOA framework and demonstrate its capacity to execute high-speed maneuvers.

\subsection{Quadrotor Model} \label{subsec: quadrotor model}
The UAV selected is a quadrotor and is modeled using standard quadrotor dynamics \cite{Beard2008} utilizing a North-East-Down (NED) coordinate system. The state vector is defined as $\boldsymbol{x}=(\boldsymbol{p}, \boldsymbol{v}, \boldsymbol{\theta}, \boldsymbol{w}_b)$, where $\boldsymbol{p}=(p_x,p_y,p_z)$ denotes the inertial position, $\boldsymbol{\theta}=(\phi,\theta,\psi)$ are the roll, pitch, and yaw Euler angles, $\boldsymbol{v}=(u,v,w)$ represents the linear velocity in the body frame, and $\boldsymbol{w}_b=(p,q,r)$ denotes the body frame angular velocity. The control input vector $\boldsymbol{u}=(\delta_f, \delta_r, \delta_b, \delta_l)$ defines the commands to the front, right, back, and left rotors, respectively, producing thrust forces $F_i = k_1\delta_i$, where $k_1$ is an experimentally determined constant.

The rigid body equations of motion are given by:
\begin{equation}\label{eq: quadrotor dynamics}
    \begin{aligned}
        \dot{\textbf{p}}& = \mathbf{R^T} \cdot \boldsymbol{v} \\
        \dot{\boldsymbol{v}} & = \frac{1}{m}\boldsymbol{F} + \mathbf{R}\cdot m\boldsymbol{g}-\boldsymbol{w}_b\times \boldsymbol{v}\\
        \dot{\boldsymbol{\theta}} & = \mathbf{G}^{-1}(\boldsymbol{\theta})\cdot \boldsymbol{w}_b \\
        \dot{\boldsymbol{w}}_b & = \mathbf{J}^{-1}(\boldsymbol{\tau} - (\boldsymbol{w}_b \times \mathbf{J}\boldsymbol{w}_b)) 
    \end{aligned}
\end{equation}
where $\mathbf{R} \in SO(3)$ is the 3-2-1 Euler rotation matrix from the vehicle to the body frame, $\mathbf{G}$ is the kinematic transformation matrix mapping Euler angle rates to body frame angular velocities, $\boldsymbol{g}$ is the gravity vector, $\mathbf{J} = \text{diag}(J_x, J_y, J_z)$ is the diagonal inertia matrix, and $m$ is the vehicle mass.

The total thrust $\boldsymbol{F}$ and control torques $\boldsymbol{\tau} = (\tau_\phi, \tau_\theta, \tau_\psi)^T$ generated by the rotors are linearly mapped from the inputs:
\begin{equation} \label{eq: quadrotor forces}
    \begin{bmatrix}
    F \\
    \tau_\phi \\
    \tau_\theta \\
    \tau_\psi
    \end{bmatrix} = \begin{bmatrix}
    k_1 & k_1 & k_1 & k_1 \\
    0 & -l k_1 & 0 & l k_1 \\
    l k_1 & 0 & -l k_1 & 0 \\
    -k_2 & k_2 & -k_2 & k_2
    \end{bmatrix}\cdot \boldsymbol{u}
\end{equation}
where $l$ is the radial arm length from the center of mass to the rotor hub, and $k_2$ relates rotor speed to yaw torque. The system parameters utilized for simulation are summarized in Table \ref{tbl: Quadrotor Parametrs}.
\begin{table}[width=.75\linewidth]
	\caption{Quadrotor parameters}
	\label{tbl: Quadrotor Parametrs}
	\renewcommand{\arraystretch}{1.3}
	\begin{tabular*}{\tblwidth}{@{\extracolsep{\fill}} c c c c @{}}
		\toprule
		Symbol & Value & Symbol & Value \\
		\midrule
		$m$ & 1.5 kg & $l$ & 0.175 m \\
		$J_x, J_y, J_z$ & 0.1 $\text{kg}\cdot \text{m}^2$ & $r_p$ & 0.05 m \\
		$k_1$ & 1.0 & $k_2$ & 1.0 \\
		\bottomrule
	\end{tabular*}
\end{table}

\section{Simulation Experiments} \label{sec: simulations}
    \subsection{Vehicle Boundary Evaluation} \label{subsec: vehicle boundary}
This subsection evaluates the computational performance of the nonconvex RCOA formulation (RNCOA) across the volumetric vehicle representations shown in Figure~\ref{fig: Volumetric Representation}. The primary objective is to quantify the computational impact of transitioning from point-geometry to multi-point vertices within RNCOA, benchmarking it against OA formulations that model a continuous vehicle boundary. The simulation features a quadrotor that must navigate around two joined rectangular obstacles (Figure~\ref{fig: scenario 1 def}).
\begin{figure}[htbp]
	\vspace{-.3cm}
	\centerline{\input{EX1_ISOVIEW_R1.pgf}} 
     \vspace{-.6cm}
    \caption{Evaluation scenario of a quadrotor navigating around two joined rectangular obstacles (dark gray). The isometric viewpoint (not to scale) illustrates the quadrotor's initial position, looking along the positive $x$-axis. The designated zones ($A_1$--$A_4$) correspond to distinct spatial navigation corridors. Inertial coordinate system $\mathcal{C}=\{x,y,z\}$ is shown at the origin, with the vehicle NED coordinate attached to the vehicle.}
    \label{fig: scenario 1 def}
\end{figure}

The designated zones, $A_1$ through $A_4$, enforce distinct topological routes (right, above, below, or left of the obstacle). These represent realistic operational scenarios where a local planner must dictate specific evasion directions while minimizing reference path error. Formulations are primarily evaluated on the computational latency required to solve the corresponding OCP. The dimensions and inertial centers of the obstacles are defined in Tables~\ref{tbl: Prism Sizing} and \ref{tbl: Ellipsoidal Sizing}. To ensure an equitable geometric comparison, baseline ellipsoidal obstacle dimensions are computed as the average of the ellipsoid inscribed within the prism and the prism inscribed within the ellipsoid.

The comparative evaluation benchmarks RNCOA against three OA formulations: the point-to-ellipsoid OA of \eqref{eq: EOA def} (EOA), the ellipsoid-to-ellipsoid OA of \eqref{eq: Ellipsoidal K} (EEOA), and the dual formulation of the minimum distance between convex sets of \eqref{eq: DMDOA def} (DMDOA). RNCOA and EOA utilize discrete point representations, EEOA models both the vehicle and obstacle as continuous ellipsoids, and DMDOA models them as strict convex sets (accommodating both point and circular vehicle footprints). For RNCOA and EOA, all area/volume representations from Figure~\ref{fig: Volumetric Representation} are simulated to assess the computational degradation as the constraint dimension increases. DMDOA can also be configured for representations in Figure~\ref{fig: Volumetric Representation}.

The exact geometric dimensions for the vehicle representations are provided in Table~\ref{tbl: Volumetric Representations}, where $l$ and $r$ denote the quadrotor arm length and rotor radius, respectively, from Table~\ref{tbl: Quadrotor Parametrs}. The standard point-geometry (PG) formulation serves as the baseline. The third column indicates the initial rotational mapping of the boundary relative to the body frame. Note that the representations of 'A' and 'C' are fixed in the vehicle frame, while 'B' and 'D' are body fixed. The quadrotor attitude is completely enveloped by representations of 'C'.

The comprehensive problem matrix is provided in Table~\ref{tbl:Ex1_problem_matrix}. For EEOA, the vehicle boundary is modeled as an equivalent ellipsoid with semi-axes matching the half-lengths in Table~\ref{tbl: Volumetric Representations}. A "default" global optimal trajectory run, where the quadrotor path is not directly constrained to $A_1$ through $A_4$, is included for all benchmark formulations.
\begin{table}[width=.9\linewidth,cols=7,pos=h]
	\caption{Prism obstacle dimensions in inertial frame}
	\renewcommand{\arraystretch}{1.3}
	\label{tbl: Prism Sizing}
	\begin{tabular*}{\tblwidth}{@{\extracolsep{\fill}} c c c c c c c @{}}
		\toprule
		& \multicolumn{3}{c}{Half-Lengths (m)} & \multicolumn{3}{c}{Center (m)} \\
		Obstacle & dx & dy & dz & x & y & z \\
		\midrule
		1 & 5 & 1 & 2 & 0 & 0 & 1 \\
		2 & 5 & 0.5 & 1 & 0 & -1.5 & 0 \\
		\bottomrule 
	\end{tabular*}
\end{table}
\vspace{-.5cm}
\begin{table}[width=.75\linewidth,pos=h]
    \centering
        \caption{Ellipsoidal obstacle dimensions in inertial frame}
        \vspace{-.2cm}
        \label{tbl: Ellipsoidal Sizing}
    \renewcommand{\arraystretch}{1.3}
    \begin{tabular*}{\tblwidth}{@{\extracolsep{\fill}} c c c c @{}}
        \toprule
        & \multicolumn{3}{ c }{Half-Lengths (m)} \\
        Obstacle & dx & dy & dz\\
        \midrule
        1 & 6.83 & 1.366 & 2.732 \\
        2 & 6.83 & 0.683 & 1.366 \\
        \bottomrule
    \end{tabular*}
\end{table}
\begin{table}[width=.9\linewidth]
	\caption{Half-lengths of the volumetric vehicle representations}
	\vspace{-.2cm}
	\renewcommand{\arraystretch}{1.5}
	\label{tbl: Volumetric Representations}
	\centering
	\begin{tabular*}{\tblwidth}{@{\extracolsep{\fill}} >{\centering\arraybackslash}p{1.8cm} >{\centering\arraybackslash}p{3.5cm} c @{}}
		\toprule
		Vehicle Representation & Dimensions (m) in Vehicle Frame & $\mathbf{R}_z$ \\
		\midrule
		PG & $[0,0,0]$ & $0^\circ$ \\
		A & $[(l+r),(l+r),0]$ & $0^\circ$ \\
		B & $[(\frac{\sqrt{2}}{2}l+r),(\frac{\sqrt{2}}{2}l+r), 0]$ & $45^\circ$\\
		C & $[(l+r),(l+r),(l+r)]$ & $0^\circ$ \\
		D & $[(\frac{\sqrt{2}}{2}l+r),(\frac{\sqrt{2}}{2}l+r),0.05]$ & $45^\circ$ \\
		\hline
	\end{tabular*}
\end{table}
\vspace{-.5cm}
\begin{table}[width=1.\linewidth]
	\centering
	\caption{Problem matrix for formulation and geometry comparison}
	\label{tbl:Ex1_problem_matrix}
	\vspace{-.2cm}
	\renewcommand{\arraystretch}{1.3}
	\begin{tabular*}{\tblwidth}{@{\extracolsep{\fill}} 
			>{\centering\arraybackslash}p{2.2cm} 
			>{\centering\arraybackslash}p{0.6cm} 
			>{\centering\arraybackslash}p{0.6cm} 
			>{\centering\arraybackslash}p{0.6cm} 
			>{\centering\arraybackslash}p{0.6cm} 
			>{\centering\arraybackslash}p{1.2cm} @{}}
		\toprule
		& \multicolumn{5}{ c }{Navigation Corridor, (Fig. \ref{fig: scenario 1 def})} \\
		OA Formulation & $A_1$ & $A_2$ & $A_3$ & $A_4$ & Default \\
		\midrule
		RNCOA &  \multicolumn{4}{ c }{PG$^{(1)}$, A, B, C, D} & NA\\
		EOA &   \multicolumn{5}{ c }{PG, A, B, C, D} \\
		EEOA &  \multicolumn{5}{ c }{C$^{(2)}$, D$^{(2)}$} \\
		DMDOA$^{(3)}$ & \multicolumn{5}{ c }{PG, A, B, C, D}\\ 
		\bottomrule
		\multicolumn{6}{l}{$^{(1)}$ \small Standard RCOA point-geometry formulation.} \\
		\multicolumn{6}{l}{$^{(2)}$ Equivalent ellipsoid inscribed within defined prism.} \\
		\multicolumn{6}{l}{$^{(3)}$ Path $A_1$ corresponds to path $A_1^*$} 
	\end{tabular*}
	\vspace{-0.5cm}
\end{table}
\vspace{1cm}

\subsubsection{Optimal Control Problem Formulation}
	
The OCP of RNCOA formulation, specifically for configuration $A_1$-D from Table~\ref{tbl:Ex1_problem_matrix}, is defined by \eqref{eq: Ex1 OCP}. The objective function minimizes three distinct penalties: the squared Euclidean reference path error along the $Y$ and $Z$ axes, the linear RCOA boundary penalty function, and a terminal state cost specifically introduced to penalize excessive vehicle deceleration. Here, the obstacle frame is the same as the world frame defined by $\mathcal{C}=\{x,y,z\}$ and let $X=x,\,Y=y,\,Z=z$.

\begin{subequations} \label{eq: Ex1 OCP}  
	\begin{align} 
		\min_{\boldsymbol{x,u,\gamma}} \sum_{k=0}^{N} \bigg[ w_1\lVert & (p_y^{(k)},p_z^{(k)}) \rVert^2_2 + w_2(\gamma_1^{(k)} + \gamma_2^{(k)})  \bigg] - w_3 p_x^{(N)}\\
		\text{s.t.} \quad \boldsymbol{x}^{(k+1)} &= f(\boldsymbol{x}^{(k)},\boldsymbol{u}^{(k)}) \label{eq: Ex1 OCP dynamics}\\
		0 &\leqslant \boldsymbol{u}^{(k)} \leqslant u_{\max} \label{eq: Ex1 OCP thrust} \\
		x_{\min}^o & \leqslant \max_{i \in \mathcal{V}} (P_{x,i}^{(k)}) + M_1 \gamma_1^{(k)} \label{eq: Ex1 OCP RCOA 1st} \\
		\min_{i \in \mathcal{V}} (P_{x,i}^{(k)}) & \leqslant x_{\max}^o + M_2 \gamma_2^{(k)}  \\
		P_{y,i}^{(k)} & \geqslant y^o_{\max} - M_3 (\gamma_1^{(k)} + \gamma_2^{(k)}) \quad \forall i \in \mathcal{V} \\
		P_{z,i}^{(k)} & \geqslant z^o_{\max} - M_4 (\gamma_1^{(k)} + \gamma_2^{(k)}) \quad \forall i \in \mathcal{V} \\
		\gamma_1^{(k)} + \gamma_2^{(k)} &\leqslant 1,\quad \gamma_1^{(k)}, \gamma_2^{(k)} \in [0,1] \label{eq: Ex1 OCP RCOA last}\\
		v_{T} & \leqslant (\mathbf{T}^{(k)})^T \boldsymbol{v}^{(k)} \label{eq: Ex1 OCP tan vel}\\
		| \psi | & \leqslant 20^\circ  \label{eq: Ex1 OCP yaw} \\ 
		\boldsymbol{P}_i^{(k)} & =\boldsymbol{p}^{(k)} + (\mathbf{R}^{(k)})^T l_{b,i}, \quad \forall i \in \{f,r,b,l\} \label{eq: Ex1 OCP boundary points}
	\end{align}
\end{subequations}
The constraint set enforces quadrotor nonlinear dynamics \eqref{eq: Ex1 OCP dynamics} defined by \eqref{eq: quadrotor dynamics}, bounded thrust limits \eqref{eq: Ex1 OCP thrust}, the RNCOA spatial exclusions \eqref{eq: Ex1 OCP RCOA 1st}--\eqref{eq: Ex1 OCP RCOA last}, a minimum tangential velocity limit \eqref{eq: Ex1 OCP tan vel}, and a bounding yaw limit for numerical stability \eqref{eq: Ex1 OCP yaw}. The kinematic spatial mapping of the vehicle's boundary vertices ($\boldsymbol{P}_i^k$) is enforced by \eqref{eq: Ex1 OCP boundary points}, where $l_{b,i}$ defines the distance from the mass center to the vertices located on the front, right, back, and left, in the body-fixed frame. The velocity constraint \eqref{eq: Ex1 OCP tan vel} ensures forward progression along the reference path, where $\mathbf{T}^{(k)}$ corresponds to the tangent vector at every $k$. The last constraint was found to improve global solver convergence across all formulations for the given time step.

While exhaustive tuning could yield performance gains, the RCOA cost weights ($w_1, w_2, w_3$) were held constant across each geometric configuration to maintain a normalized baseline. They were configured such that no obstacle collision occurred in any simulation.

To constrain the benchmark formulations (EOA, EEOA, DMDOA) into designated topological corridors (e.g., path $A_1$), hard spatial quadrant restrictions and corresponding initial guesses are required. For example, to constrain the path to $A_1$ requires the following additional constraints in the vehicle frame:
\begin{equation*}
	\begin{aligned}
		p_y^{(k)} & \geqslant -0.10 \\
		p_z^{(k)} & \geqslant -0.25 \\
		p_{y,\text{initial}}^{(k)} & = 2.0 \\
		p_{z,\text{initial}}^{(k)} & = 1.0
	\end{aligned}
\end{equation*}

\subsubsection{Evaluation Results}
The array of OCPs were transcribed using CasADi \cite{Andersson2018}, employing a multiple-shooting scheme with an RK4 integrator. The discrete temporal horizon was fixed at $N=80$ nodes. The total look-ahead simulation time varied slightly among formulations (Table~\ref{tbl: Ex1_Simulation_time}); these variations were strictly mandated by solver convergence, as EEOA and DMDOA routinely failed to converge at the aggressive time steps achievable by RNCOA and EOA. All trajectories shared identical initial origins, varying only in initial forward velocity $u$ as detailed in Table~\ref{tbl: Ex1_ICs}. The primary reference path was aligned precisely along the inertial $X$-axis.

\begin{table}[width=.75\linewidth,cols=2,pos=h]
	\caption{Simulation horizon time ($T_p$ sec.) per OA}
	\label{tbl: Ex1_Simulation_time}
	\renewcommand{\arraystretch}{1.3}
	\begin{tabular*}{\tblwidth}{@{} LL@{} }
		\toprule
		RNCOA, EOA & 4.8 \\
		EEOA & 3.5 \\
		DMDOA & 3.75 \\
		\bottomrule
	\end{tabular*}
\end{table}

\begin{table}[width=.5\linewidth,cols=2,pos=h]
	\caption{Initial state conditions, $\boldsymbol{x}_0$}
	\label{tbl: Ex1_ICs}
	\renewcommand{\arraystretch}{1.3}
	\begin{tabular*}{\tblwidth}{@{} LL@{} }
		\toprule
		Variable & Value \\
		\midrule
		$\boldsymbol{p}$ & [-15, 0, 0] (m) \\
		$u$ & 12 or 8* (m/s) \\
		\bottomrule
	\end{tabular*}
	
	\vspace{2pt}
	\small $^*$ Exclusively applied for $A_2$
\end{table}

\begin{figure*}[!] 
	\centerline{\input{Ex1_summary.pgf}} 
	\caption{Box plot computational latency summary for solving the OA OCPs in Table~\ref{tbl:Ex1_problem_matrix}. The distributions represent 20 independent solver executions across all designated topological corridors ($A_1$--$A_4$) and vehicle representations (including the default PG trajectory). RNCOA consistently demonstrates the lowest median solver time and highest consistency (narrowest interquartile range) across all tested configurations.}
	\label{fig: Ex1 Summary Box Plot}
\end{figure*}

RNCOA, EOA, and EEOA OCPs were solved with FATROP \cite{Vanroye2023}. Due to structural matrix complexities, DMDOA required IPOPT \cite{Wachter2006} with the HSL MA57 linear solver \cite{Duff2004}, frequently relying on adaptive or probing barrier strategies to achieve convergence. Both solvers were strictly bound to a $10^{-6}$ error tolerance. Benchmarking was performed on an HP OMEN desktop (Intel i7-14700F CPU, 32GB RAM).

Since all formulations yielded collision-free paths (excluding inter-sample intersections), computational latency serves as the primary performance metric. Figure~\ref{fig: Ex1 Summary Box Plot} summarizes the statistical distributions of 20 independent executions per configuration. Across all topological corridors, RNCOA consistently achieved the lowest median computation times and demonstrated unparalleled consistency via the narrowest interquartile ranges. EOA ranked second, followed by EEOA, while DMDOA exhibited the highest latency variance. Although DMDOA and EEOA mathematically guarantee continuous 3D non-intersection, this geometric fidelity imposes a prohibitive computational penalty that undermines their viability for real-time control.

Table~\ref{tbl: Ex1 Average RNCOA and EOA} compares the mean run times for RNCOA and EOA across the four vehicle representations against the baseline point-geometry (PG) configuration. While all volumetric EOA configurations incur higher latency than their PG baseline, RNCOA computationally outperforms its own PG reference in every configuration except 'D'.
\begin{table}[h]
	\centering
	\caption{Mean Vehicle Representation Performance (sec.)}
	\vspace{-.2cm}
	\label{tbl: Ex1 Average RNCOA and EOA}
	\renewcommand{\arraystretch}{1.3}
	\begin{tabular*}{\tblwidth}{@{} LLLLLL@{} }
		\toprule[1pt] 
		OA/Config. & A & B & C & D & Ref (PG)\\ 
		\midrule
		RNCOA & 0.3597 & 0.3194 & 0.3579 & 0.4169 & 0.3874\\ 
		EOA   & 0.5097 & 0.4584 & 0.4677 & 0.5935 & 0.3797\\ \bottomrule[1pt]
	\end{tabular*}
\end{table}

\begin{figure}[htbp]
	\vspace{-1.5em} 
	\centerline{\includegraphics[width=\columnwidth]{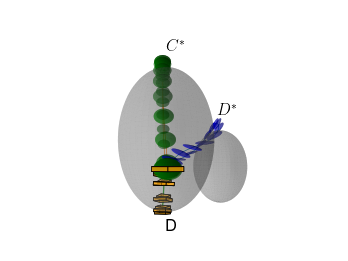}}
	\vspace{-.6cm}
	\caption{Locally optimal 3D trajectories generated by EEOA and EOA baselines. Highlighted: Configuration C* (bounding sphere) navigating path $A_2$ (green), Configuration D* (bounding ellipsoid) traversing $A_1$ (blue), and EOA Configuration D maneuvering through $A_4$ (orange). View is along the positive $x$-axis.}
	\label{fig: Ex1_EEOA_trajectories}
\end{figure}

Selected topological trajectories are visualized in Figures~\ref{fig: Ex1_RNCOA_trajectories} and \ref{fig: Ex1_EEOA_trajectories}. Figure~\ref{fig: Ex1_RNCOA_trajectories} displays the RNCOA trajectories for routes $A_1-D$, $A_2-C$, and $A_4-B$. Figure~\ref{fig: Ex1_EEOA_trajectories} overlays the equivalent spatial solutions utilizing the benchmark ellipsoidal obstacle configurations. The advantage of a tightly coupled body-frame representation over statically enlarged bounds is visually evident: the quadrotor can be seen executing high-performance banking maneuvers to navigate tightly around the obstacle vertices.

\begin{figure}[htbp] 
	\centerline{\input{EX1_Figure_RCOA.pgf}} 
	\vspace{-.5cm}
	\caption{Locally optimal 3D trajectories generated by RNCOA. Highlighted: Configuration C navigating path $A_2$ (green), Configuration D traversing $A_1$ (blue), and Configuration B maneuvering through $A_4$ (orange).}
	\label{fig: Ex1_RNCOA_trajectories}
\end{figure}

    \subsection{Narrow Passage}
In this simulation, the prediction horizon is aggressively reduced to demonstrate two critical capabilities of the RNCOA framework: (1) its ability to successfully guide the vehicle even when the obstacle resides strictly outside the finite prediction horizon, and (2) its capacity to significantly reduce computational latency for high-frequency, real-time NMPC deployment. The scenario requires the quadrotor to navigate through a narrow, restrictive slit, a maneuver that demands a high roll angle to maintain geometric feasibility.

The passage is defined with a width of $0.25$~m and a depth of $1.0$~m. For context, the physical radius of the quadrotor (from the center of mass to the outer rotor tip) is $0.225$~m. For geometric fidelity, the vehicle is modeled using the rotating 2D planar boundary ('B' from Table~\ref{tbl: Volumetric Representations}).

The quadrotor is controlled via a dual-mode NMPC framework. The primary OCP actively manages obstacle avoidance and trajectory tracking. Because of the aggressive maneuver, a secondary, computationally lightweight OCP commanding hover is activated immediately after the vehicle clears the passage, similar to other work \cite{Mellinger2014}. The performance of RNCOA in this extreme scenario is benchmarked against the EOA formulation.

\subsubsection{Optimal Control Problem Formulation}

The primary OA OCP is formalized in \eqref{eq: Ex2 OCP}. To facilitate the aggressive maneuver, the maximum allowable thrust constraint \eqref{eq: Ex2 OCP thrust} is significantly relaxed compared to the previous simulation. The cost function employs an $L_1$ norm penalty on the lateral ($p_y$) and vertical ($p_z$) path deviations to prioritize exact reference tracking without overly penalizing transient aggressiveness.

To geometrically define the "slit," the RCOA constraints are symmetrically duplicated. The first set \eqref{eq: Ex2 OCP RCOA 1st}--\eqref{eq: Ex2 OCP o1 zmax} defines the lower-left boundary (forcing the quadrotor above and to the right), while the second set defines the top-right boundary (forcing the quadrotor below and to the left). Together, they form a strict traversable corridor.

The OCP tuning parameters are summarized in Table~\ref{tbl: Ex2 OCP parameters}. The reference trajectory is configured as a linear path parallel to the inertial $X$-axis, rendering $r_y$ and $r_z$ constant. The secondary stabilizing OCP is functionally straightforward, structured purely to penalize linear velocities to enforce a zero-velocity hover state.

\begin{subequations} 
	\label{eq: Ex2 OCP}  
	\begin{align} 
		\min_{\boldsymbol{x,u,\gamma}} \sum_{k=0}^{N} \bigg[ w_1\lVert & p_y^{(k)}-r_y \rVert_1 +w_2\lVert p_z^{(k)}-r_z\rVert_1 +w_3 \sum_{j=1}^4 \gamma_j^{(k)} \bigg] \\
		\text{s.t.} \quad \boldsymbol{x}^{(k+1)} &= f(\boldsymbol{x}^{(k)},\boldsymbol{u}^{(k)}) \label{eq: Ex2 OCP dynamics}\\
		0 &\leqslant \boldsymbol{u}^{(k)} \leqslant 20\,\text{N} \label{eq: Ex2 OCP thrust} \\
		x_{\min}^{o_1} & \leqslant \max_{i \in \mathcal{V}} (P_{x,i}^{(k)}) +M_1 \gamma_1^{(k)} \label{eq: Ex2 OCP RCOA 1st} \\
		\min_{i \in \mathcal{V}} (P_{x,i}^{(k)}) & \leqslant x_{\max}^{o_1} + M_2 \gamma_2^{(k)}  \\
		P_{y,i}^{(k)} & \geqslant y^{o_1}_{\max} - M_3 (\gamma_1^{(k)} + \gamma_2^{(k)}) \quad \forall i \in \mathcal{V} \label{eq: Ex2 OCP o1 ymax}\\
		P_{z,i}^{(k)} & \geqslant z^{o_1}_{\max} - M_4 (\gamma_1^{(k)} + \gamma_2^{(k)}) \quad \forall i \in \mathcal{V} \label{eq: Ex2 OCP o1 zmax}\\
		x_{\min}^{o_2} & \leqslant \max_{i \in \mathcal{V}} (P_{x,i}^{(k)}) +M_1 \gamma_3^{(k)} \\
		\min_{i \in \mathcal{V}} (P_{x,i}^{(k)}) & \leqslant x_{\max}^{o_2} + M_2 \gamma_4^{(k)}  \\
		P_{y,i}^{(k)} & \leqslant y^{o_2}_{\min} + M_3 (\gamma_3^{(k)} + \gamma_4^{(k)}) \quad \forall i \in \mathcal{V} \\
		P_{z,i}^{(k)} & \leqslant z^{o_2}_{\min} + M_4 (\gamma_3^{(k)} + \gamma_4^{(k)}) \quad \forall i \in \mathcal{V} \\
		\sum_{j=1}^2 \gamma_j^{(k)}&\leqslant 1,\;\sum_{j=3}^4 \gamma_j^{(k)} \leqslant 1,\quad \gamma_j^{(k)} \in [0,1] \label{eq: Ex2 OCP RCOA last}\\
		\boldsymbol{P}_i^{(k)} & =\boldsymbol{p}^{(k)} + (\mathbf{R}^{(k)})^T l_{b,i}, \; \forall i \in \{f,r,b,l\} \label{eq: Ex2 OCP boundary points}
	\end{align}
\end{subequations}
\begin{table}[width=.9\linewidth]
	\caption{Narrow Passage OCP Parameters}
	\label{tbl: Ex2 OCP parameters}
	\renewcommand{\arraystretch}{1.3}
	\begin{tabular*}{\tblwidth}{@{\extracolsep{\fill}} c c c c @{}}
		\toprule
		Parameter & Value & Parameter & Value \\
		\midrule
		$w_1,\,w_2,\,w_3$ & 18, 0.05, 190 & $r_y,\,r_z$ & 1.1, 2.0 (m) \\
		$x^{o_*}_{min},\,x^{o_*}_{max}$ & -0.1, 1.0 (m) & $y^{o_1}_{\max},\,y^{o_2}_{\min}$ & 1.0, 1.25 (m) \\
		$z^{o_1}_{max},\,z^{o_2}_{min}$ & 1.0, 3.0 (m) & $M_{1-4}$ & 50, 50, 3, 3 \\
		\bottomrule
	\end{tabular*}
\end{table}

    \subsubsection{Narrow Passage Results}
    
To evaluate the RNCOA formulation, and equivalently RCOA, under severe prediction horizon constraints, the horizon was heavily truncated to $T_p=0.25$~s, utilizing a mere $N = 9$ temporal nodes. This configuration maximizes the achievable control update frequency. The simulation relied on an NMPC controller, for a total duration of $2.5s$, where each OCP was solved using FATROP \cite{Vanroye2023}. To further prioritize real-time controller response, the solver tolerance was relaxed to $10^{-1}$. Initial state conditions are detailed in Table~\ref{tbl: Ex2_ICs}, with all unlisted states initialized to zero.  

\begin{table}[width=.5\linewidth]
    \centering
    \caption{Initial conditions, $\boldsymbol{x}_0$}
    \vspace{-.2cm}
    \label{tbl: Ex2_ICs}
    \renewcommand{\arraystretch}{1.3}
    \begin{tabular*}{\tblwidth}{@{\extracolsep{\fill}} c c @{}}
        \toprule
        Variable & Value \\
        \midrule
        $\boldsymbol{p}$ & [-10, 1.1, 2.0] (m) \\
        $u$ & 12 (m/s) \\
        \bottomrule
    \end{tabular*}
\end{table}
The closed-loop trajectory is illustrated in Figures~\ref{fig: Ex2_trajectory} and \ref{fig: Ex2_MPC_iterations}. As shown in Figure~\ref{fig: Ex2_trajectory}, the quadrotor approaches the passage with zero initial roll, executing an aggressive, high-angle roll maneuver strictly to conform to the narrow geometry of the slit without collision.

To illustrate the receding horizon behavior during this approach, Figure~\ref{fig: Ex2_MPC_iterations} visualizes multiple NMPC solution sets (i.e., $\phi(i;\boldsymbol{x}^*,\boldsymbol{u}^*),\;i\in\{16,19,22\}$), explicitly rendered as three overlaid solution sets highlighted in blue, yellow, and orange. Notably, the first two (blue and yellow) open-loop prediction horizons terminate strictly before the vehicle reaches the physical passage. Despite the obstacle remaining outside the solver's look-ahead window in these two iterations, each distinct sequence correctly actuates the required evasive roll angle. This visual evidence strongly confirms the theoretical assertion of Equation \eqref{eq: case 3 opt YZ}: the continuous linear relation acts as an anticipatory gradient, inherently regulating the optimal attitude well in advance of the hard spatial boundary.

Finally, upon exiting the passage, the secondary OCP successfully arrests the resulting dynamic instability, guiding the vehicle through a brief helical dissipation path into a stable hover.

Conversely, the EOA formulation failed under identical NMPC configurations. To mitigate numerical drift, the EOA solver tolerance was tightened to $10^{-3}$. As illustrated by the orange boundaries in Figures~\ref{fig: Ex2_EOA_trajectory} and \ref{fig: Ex2_EOA_trajectory2}, the simulation terminated prematurely when the solver declared the OCP infeasible. The final valid NMPC solution set before termination is rendered in blue. For visual clarity, the top and bottom obstacles encapsulating the slit are omitted from these figures. Note that while the bounding ellipsoids increase the effective depth of the obstacles, they are sized to tangentially meet the target slit boundary (black outline), as depicted in Figure~\ref{fig: Ex2_EOA_trajectory2}.

By the time the finite horizon was long enough to "see" the bounding ellipsoid, the vehicle's momentum rendered the required evasion physically impossible, resulting in an infeasible trajectory state. (Note: to ensure a fair kinematic baseline, a minimum velocity constraint of $8$~m/s was enforced on the EOA trial to prevent artificial deceleration outside the passage).

\begin{figure}[htbp] 
	\centering
	\resizebox{0.75\columnwidth}{0.23\textheight}{\input{EX2_narrowpas_trajectory.pgf}}
	\vspace{-.2cm}
	\caption{Closed-loop RNCOA trajectory navigating a narrow slit. A 2D planar outline of the quadrotor body illustrates the strict rotational conformity required to maintain geometric feasibility through the passage.}
	\label{fig: Ex2_trajectory}
\end{figure}
\begin{figure}[htbp] 
	\centerline{\input{Ex3_NS_multiIter_R2.pgf}} 
	\vspace{-.2cm}
	\caption{Sequential NMPC prediction horizons demonstrating receding horizon behavior. Three distinct OCP solution sets (highlighted in blue, yellow, and orange) are overlaid. Notably, the blue and yellow finite prediction horizons terminate strictly prior to intersection with the passage, highlighting the RCOA formulation's capacity to induce anticipatory optimal rotations independently of the look-ahead boundary.}
	\label{fig: Ex2_MPC_iterations}
\end{figure}
Beyond achieving task feasibility, RNCOA demonstrated exceptional numerical efficiency. The computational statistics for the OA OCP during the simulated maneuver are provided in Table~\ref{tbl: Ex2_solver_stats}. RNCOA maintained an average solve time of $11.3$~ms, firmly enabling sustained control frequencies exceeding $30$~Hz.

\begin{table}[width=0.7\linewidth]
    \centering
    \caption{OA OCP solver performance (sec.)}
    \vspace{-.2cm}
    \label{tbl: Ex2_solver_stats}
    \renewcommand{\arraystretch}{1.3}
	\begin{tabular*}{\tblwidth}{@{\extracolsep{\fill}} c c c c @{}}
		\toprule[1pt] 
		OA & mean & max & min \\ 
		\midrule
		RNCOA & 0.0113 & 0.0162 & 0.0075 \\ 
		EOA   & 0.0168 & 0.0233 & 0.0135 \\ 
		\bottomrule[1pt]
	\end{tabular*}
\end{table}

\begin{figure}[htbp]
	\centering
	\vspace{-1.em}
	\centerline{\input{Ex3_EOA_1_R1.pgf}} 
	\vspace{-0.8cm}
	\caption{Failed NMPC trajectory employing the EOA formulation under a truncated prediction horizon. The executed trajectory (orange) leads to solver infeasibility strictly prior to passage entry. Because EOA relies on hard spatial constraints without an anticipatory gradient, the controller remains myopic; the vehicle's momentum renders evasion physically impossible by the time the obstacle enters the finite prediction window (blue).}
	\label{fig: Ex2_EOA_trajectory}
\end{figure}

\begin{figure}[htbp] 
	\centering
	\vspace{-2.25em}
	\centerline{\input{Ex3_EOA_2_R1.pgf}} 
	\vspace{-1.1cm}
	\caption{Orthogonal view along the $+X$-direction detailing the EOA terminal failure state. This perspective highlights the severe spatial conflict caused by horizon blindness. Lacking the early regulatory guidance provided by RCOA, the EOA controller fails to initiate the required high-angle roll maneuver in advance, driving the vehicle into a kinematically irrecoverable state.}
	\label{fig: Ex2_EOA_trajectory2}
\end{figure}

\subsection{Discussion}
In its foundational 2D introduction \cite{Tapia2025}, RCOA demonstrated computational performance on par with, or superior to, established OA formulations within nonconvex domains (e.g., bicycle models with nonlinear tire dynamics). That performance was particularly notable in tightly constrained environments where geometrically feasible trajectories were exceedingly narrow. The 3D and volumetric extensions introduced in this paper confirm that this computational advantage scales effectively to higher-dimensional UAV dynamics.

The introduction of the nonconvex multi-point OA formulation (RNCOA) was specifically designed to mitigate excessive constraints and variables that would otherwise be required by RCOA. The statistical results (Figure~\ref{fig: Ex1 Summary Box Plot}) confirm that RNCOA exhibits unparalleled consistency and the lowest median latency among the tested methods. While this discrete multi-point approximation sacrifices the absolute, continuous geometric guarantees of exact 3D formulations (such as EEOA or DMDOA, which mathematically bound the entire continuum of the vehicle volume), the results (Table~\ref{tbl: Ex1 Average RNCOA and EOA}) indicate that the number of discrete vertices used to approximate the vehicle can be significantly increased while still maintaining a dominant computational advantage.

A critical observation regarding trajectory quality lies in the inter-sample behavior of the formulations. As highlighted in the github repository \href{https://rtapia8484.github.io/RCOA/}{rtapia8484.github.io/RCOA}, a direct side-by-side comparison of RNCOA and DMDOA along path $A_2-D$  reveals stark differences in spatial enforcement. Because DMDOA enforces hard constraints strictly at the discrete temporal nodes of the prediction horizon, the vehicle trajectory is susceptible to severe inter-sample constraint violations (e.g., corner clipping between nodes).  Conversely, RNCOA natively adheres to the regulatory gradient described by \eqref{eq: case 3 opt YZ}. As soon as a single boundary vertex triggers the spatial condition, the analytical penalty naturally regulates the entire boundary set safely above $z^o_{\max}$. This intrinsic buffering effectively eliminates inter-sample collision without requiring infinitely dense node spacing.

Furthermore, RNCOA successfully preserves the unique characteristic of its 2D predecessor: the capacity to command optimal evasive maneuvers even when the obstacle resides outside the finite prediction horizon. While operating with severely truncated horizons technically eliminates the solver's ability to formally certify trajectory feasibility in advance, this limitation can be addressed architecturally. A dual horizon NMPC framework can be employed to guarantee strict safety; a slow-rate, long horizon NMPC can evaluate global feasibility, while a high-rate, short horizon RNCOA controller (executing at the $11.3$~ms median latency shown in Table~\ref{tbl: Ex2_solver_stats}) tracks the state and dynamically executes the avoidance maneuver at control frequencies safely exceeding $30$~Hz.

\section{Conclusion and Future Work} \label{sec: Conclusion}
This paper presented the three-dimensional extension of RCOA and applied it to the computationally demanding problem of UAV volumetric navigation. By introducing a nonconvex multi-point geometry modification (RNCOA), the framework allows for high-fidelity, rotating vehicle footprints to navigate tightly constrained environments without the conservatism of obstacle inflation required by point-geometry approximations.

Extensive numerical simulations demonstrate that RNCOA significantly outperforms baseline formulations, including ellipsoidal and the dual problem of the minimum distance constraints, in both raw computational speed and solver consistency. Most notably, the formulation's KKT-derived continuous penalty gradient enables the NMPC to inherently anticipate obstacles and execute aggressive evasive maneuvers independently of the finite prediction horizon length. This attribute breaks the traditional compromise between controller look-ahead capability and real-time execution frequency.

A structural requirement of RCOA is the necessity to define parallel OCPs to explore distinct topological routing options. While this scales the problem, these OCPs are entirely decoupled and can be solved simultaneously. In practical deployment, this aligns seamlessly with hierarchical planning architectures, where a global behavioral planner \cite{Ziegler2014} dictates the preferred spatial homotopy, or the use of an RRT-type algorithm, isolating a single optimal route for the RNCOA NMPC to execute.

Future work will focus on generalizing the analytical properties of \eqref{eq: RCOA OPT RESULT} to design customized penalty formulations that induce arbitrary optimal behavioral responses beyond standard Euclidean distance regulation. Additionally, the underlying relaxed indicator logic of RCOA presents promising avenues for integration into broader Mixed Logical Dynamical (MLD) and hybrid control systems, potentially circumventing the need for computationally prohibitive MIP in state-dependent logic transitions.

\printcredits
\bibliography{export}

@article{SCHOELS20206555,
	title = {CIAO⁎: MPC-based Safe Motion Planning in Predictable Dynamic Environments},
	journal = {IFAC-PapersOnLine},
	volume = {53},
	number = {2},
	pages = {6555-6562},
	year = {2020},
	note = {21st IFAC World Congress},
	issn = {2405-8963},
	doi = {https://doi.org/10.1016/j.ifacol.2020.12.072},
	url = {https://www.sciencedirect.com/science/article/pii/S2405896320303281},
	author = {Tobias Schoels and Per Rutquist and Luigi Palmieri and Andrea Zanelli and Kai O. Arras and Moritz Diehl}
}

@misc{Tapia2025,
      title={A Convex Obstacle Avoidance Formulation}, 
      author={Ricardo Tapia and Iman Soltani},
      year={2025},
      eprint={2512.13836},
      archivePrefix={arXiv},
      primaryClass={eess.SY},
      url={https://arxiv.org/abs/2512.13836}, 
}

@inproceedings{Ziegler2014,
	author = {J Ziegler and P Bender and T Dang and C Stiller},
	doi = {10.1109/IVS.2014.6856581},
	isbn = {1931-0587},
	booktitle = {2014 IEEE Intelligent Vehicles Symposium Proceedings},
	pages = {450-457},
	title = {Trajectory planning for Bertha — A local, continuous method},
	year = {2014}
}

@article{Allen2019,
	author = {Ross E Allen and Marco Pavone},
	doi = {https://doi.org/10.1016/j.robot.2018.11.017},
	issn = {0921-8890},
	journal = {Robotics and Autonomous Systems},
	pages = {174-193},
	title = {A real-time framework for kinodynamic planning in dynamic environments with application to quadrotor obstacle avoidance},
	volume = {115},
	url = {https://www.sciencedirect.com/science/article/pii/S0921889017308692},
	year = {2019}
}

@article{Han2025,
	author = {Qiang Han and Xingyuan Ma and Jilong Liu and Hanlin Liu and Yuehao Yan and Qianguo Yang},
	doi = {10.1038/s41598-025-32993-w},
	issn = {2045-2322},
	issue = {1},
	journal = {Scientific Reports},
	pages = {3089},
	title = {A hybrid RRT-DWA path planning framework for UAVs in dynamic environments},
	volume = {16},
	url = {https://doi.org/10.1038/s41598-025-32993-w},
	year = {2025}
}

@article{SuryaPrakash2025,
	author = {S K Surya Prakash and Darshankumar Prajapati and Bhuvan Narula and Amit Shukla},
	doi = {10.3389/frobt.2025.1604506},
	issn = {2296-9144},
	journal = {Frontiers in Robotics and AI},
	title = {iAPF: an improved artificial potential field framework for asymmetric dual-arm manipulation with real-time inter-arm collision avoidance},
	volume = {Volume 12 - 2025},
	url = {https://www.frontiersin.org/journals/robotics-and-ai/articles/10.3389/frobt.2025.1604506},
	year = {2025}
}

@article{Dobrevski2024,
	author = {Matej Dobrevski and Danijel Skočaj},
	doi = {10.1109/TRO.2024.3400932},
	journal = {IEEE Transactions on Robotics},
	pages = {3068-3081},
	title = {Dynamic Adaptive Dynamic Window Approach},
	volume = {40},
	year = {2024}
}

@inproceedings{Perez2012,
	author = {Alejandro Perez and Robert Platt and George Konidaris and Leslie Kaelbling and Tomas Lozano-Perez},
	doi = {10.1109/ICRA.2012.6225177},
	booktitle = {2012 IEEE International Conference on Robotics and Automation},
	pages = {2537-2542},
	title = {LQR-RRT*: Optimal sampling-based motion planning with automatically derived extension heuristics},
	year = {2012}
}

@article{Wachter2006,
   author = {W{\"{a}}chter, Andreas and Biegler, Lorenz T},
   doi = {10.1007/s10107-004-0559-y},
   issn = {1436-4646},
   issue = {1},
   journal = {Mathematical Programming},
   pages = {25-57},
   title = {On the implementation of an interior-point filter line-search algorithm for large-scale nonlinear programming},
   volume = {106},
   url = {https://doi.org/10.1007/s10107-004-0559-y},
   year = {2006}
}

@misc{Vanroye2023,
   author = {Lander Vanroye and Ajay Sathya and Joris De Schutter and Wilm Decré},
   title = {FATROP : A Fast Constrained Optimal Control Problem Solver for Robot Trajectory Optimization and Control},
   url = {https://arxiv.org/abs/2303.16746},
   year = {2023}
}

@article{Andersson2018,
   author = {Joel A E Andersson and Joris Gillis and Greg Horn
and James B Rawlings and Moritz Diehl},
   journal = {Mathematical Programming Computation},
   title = {CasADi – A software framework for nonlinear optimization
and optimal control},
   year = {2018}
}

@inproceedings{Beard2008,
   author = {Randal W Beard},
   title = {Quadrotor Dynamics and Control},
   url = {https://api.semanticscholar.org/CorpusID:195351003},
   year = {2008}
}

@article{Bemporad1999ControlConstraints,
    title = {Control of systems integrating logic, dynamics, and constraints},
    year = {1999},
    journal = {Automatica},
    author = {Bemporad, Alberto and Morari, Manfred},
    pages = {407--427},
    volume = {35}
}

@inproceedings{Gilitschenski2014,
   author = {Igor Gilitschenski and Uwe D Hanebeck},
   doi = {10.1109/SDF.2014.6954724},
   booktitle = {2014 Sensor Data Fusion: Trends, Solutions, Applications (SDF)},
   pages = {1-6},
   title = {A direct method for checking overlap of two hyperellipsoids},
   year = {2014}
}

@book{Boyd2014ConvexOptimization,
    title = {{Convex Optimization}},
    year = {2014},
    author = {Boyd, Stephen P and Vandenberghe, Lieven},
    publisher = {Cambridge University Press},
    url = {https://web.stanford.edu/%7Eboyd/cvxbook/},
    isbn = {978-0-521-83378-3},
    doi = {10.1017/CBO9780511804441}
}

@article{Pereira2021,
   author = {Jean C Pereira and Valter J S Leite and Guilherme V Raffo},
   doi = {10.1007/s10846-021-01310-8},
   issn = {1573-0409},
   issue = {3},
   journal = {Journal of Intelligent \& Robotic Systems},
   pages = {62},
   title = {Nonlinear Model Predictive Control on SE(3) for Quadrotor Aggressive Maneuvers},
   volume = {101},
   url = {https://doi.org/10.1007/s10846-021-01310-8},
   year = {2021}
}

@inproceedings{Tracy2023,
   author = {Kevin Tracy and Taylor A Howell and Zachary Manchester},
   doi = {10.1109/ICRA48891.2023.10160716},
   booktitle = {2023 IEEE International Conference on Robotics and Automation (ICRA)},
   pages = {3663-3670},
   title = {Differentiable Collision Detection for a Set of Convex Primitives},
   year = {2023}
}

@misc{Garg2025,
   author = {Shruti Garg and Thomas Cohn and Russ Tedrake},
   title = {Planning Shorter Paths in Graphs of Convex Sets by Undistorting Parametrized Configuration Spaces},
   url = {https://arxiv.org/abs/2411.18913},
   year = {2025}
}

@misc{Morozov2024,
   author = {Savva Morozov and Tobia Marcucci and Alexandre Amice and Bernhard Paus Graesdal and Rohan Bosworth and Pablo A Parrilo and Russ Tedrake},
   title = {Multi-Query Shortest-Path Problem in Graphs of Convex Sets},
   url = {https://arxiv.org/abs/2409.19543},
   year = {2024}
}

@article{She2025,
   author = {Yang She and Chao Song and Zetian Sun and Bo Li},
   doi = {10.3390/drones9010039},
   issn = {2504-446X},
   issue = {1},
   journal = {Drones},
   title = {Optimized Model Predictive Control-Based Path Planning for Multiple Wheeled Mobile Robots in Uncertain Environments},
   volume = {9},
   url = {https://www.mdpi.com/2504-446X/9/1/39},
   year = {2025}
}

@inproceedings{LiuXiaofengandChen2018,
   author = {Hailin
and Wang Chengcheng
and Hu Fang
and Yang Xianqiang Liu Xiaofeng
and Chen},
   city = {Cham},
   editor = {Alexandre
and Yan Yamin
and Chen Shifeng Chen Zhiyong
and Mendes},
   isbn = {978-3-319-97589-4},
   booktitle = {Intelligent Robotics and Applications},
   pages = {170-181},
   publisher = {Springer International Publishing},
   title = {MPC Control and Path Planning of Omni-Directional Mobile Robot with Potential Field Method},
   year = {2018}
}

@article{Gilbert1988,
   author = {E G Gilbert and D W Johnson and S S Keerthi},
   doi = {10.1109/56.2083},
   issue = {2},
   journal = {IEEE Journal on Robotics and Automation},
   pages = {193-203},
   title = {A fast procedure for computing the distance between complex objects in three-dimensional space},
   volume = {4},
   year = {1988}
}

@inproceedings{Gilbert1994,
   author = {E G Gilbert and Chong Jin Ong},
   doi = {10.1109/ROBOT.1994.351237},
   booktitle = {Proceedings of the 1994 IEEE International Conference on Robotics and Automation},
   pages = {579-586 vol.1},
   title = {New distances for the separation and penetration of objects},
   year = {1994}
}

@misc{Leprich2025,
   author = {David Leprich and Mario Rosenfelder and Markus Herrmann-Wicklmayr and Kathrin Flaßkamp and Peter Eberhard and Henrik Ebel},
   title = {Efficient Collision-Avoidance Constraints for Ellipsoidal Obstacles in Optimal Control: Application to Path-Following MPC and UAVs},
   url = {https://arxiv.org/abs/2510.26531},
   year = {2025}
}

@article{Ros2002,
   author = {L Ros and A Sabater and F Thomas},
   doi = {10.1109/TSMCB.2002.1018763},
   issue = {4},
   journal = {IEEE Transactions on Systems, Man, and Cybernetics, Part B (Cybernetics)},
   pages = {430-442},
   title = {An ellipsoidal calculus based on propagation and fusion},
   volume = {32},
   year = {2002}
}

@article{Rosenfelder2025,
   author = {Mario Rosenfelder and Hendrik Carius and Markus Herrmann-Wicklmayr and Peter Eberhard and Kathrin Flaßkamp and Henrik Ebel},
   doi = {10.1016/J.MECHATRONICS.2025.103386},
   issn = {0957-4158},
   journal = {Mechatronics},
   month = {10},
   pages = {103386},
   publisher = {Pergamon},
   title = {Efficient avoidance of ellipsoidal obstacles with model predictive control for mobile robots and vehicles},
   volume = {110},
   url = {https://www.sciencedirect.com/science/article/pii/S0957415825000959},
   year = {2025}
}

@article{Duff2004,
   author = {Iain S. Duff},
   doi = {10.1145/992200.992202},
   issn = {00983500},
   issue = {2},
   journal = {ACM Transactions on Mathematical Software},
   title = {MA57 - A code for the solution of sparse symmetric definite and indefinite systems},
   volume = {30},
   year = {2004}
}

@inproceedings{Chen2017,
   author = {Jianyu Chen and Wei Zhan and Masayoshi Tomizuka},
   doi = {10.1109/ITSC.2017.8317745},
   booktitle = {2017 IEEE 20th International Conference on Intelligent Transportation Systems (ITSC)},
   pages = {1-7},
   title = {Constrained iterative LQR for on-road autonomous driving motion planning},
   year = {2017}
}

@article{Zhang2021,
   author = {Xiaojing Zhang and Alexander Liniger and Francesco Borrelli},
   doi = {10.1109/TCST.2019.2949540},
   issue = {3},
   journal = {IEEE Transactions on Control Systems Technology},
   pages = {972-983},
   title = {Optimization-Based Collision Avoidance},
   volume = {29},
   year = {2021}
}

@article{Romero2022,
   author = {Angel Romero and Sihao Sun and Philipp Foehn and Davide Scaramuzza},
   doi = {10.1109/TRO.2022.3173711},
   issn = {19410468},
   issue = {6},
   journal = {IEEE Transactions on Robotics},
   title = {Model Predictive Contouring Control for Time-Optimal Quadrotor Flight},
   volume = {38},
   year = {2022}
}

@inproceedings{Schouwenaars2001,
   author = {Tom Schouwenaars and Bart De Moor and Eric Feron and Jonathan How},
   doi = {10.23919/ecc.2001.7076321},
   isbn = {9783952417362},
   booktitle = {2001 European Control Conference, ECC 2001},
   pages = {2603-2608},
   publisher = {Institute of Electrical and Electronics Engineers Inc.},
   title = {Mixed integer programming for multi-vehicle path planning},
   year = {2001}
}

@book{Khatib1985,
   author = {Oussama Khatib},
   doi = {10.1109/ROBOT.1985.1087247},
   month = {4},
   pages = {500-505},
   title = {Real-Time Obstacle Avoidance for Manipulators and Mobile Robots},
   volume = {1},
   year = {1985}
}

@book{Borenstein1990,
   author = {J Borenstein and Yoram Koren},
   doi = {10.1109/ROBOT.1990.126042},
   isbn = {0-8186-9061-5},
   month = {6},
   pages = {572-577 vol.1},
   title = {Real-time obstacle avoidance for fast mobile robots in cluttered environments},
   year = {1990}
}

\bio{Headshot_R1}{Tapia, Ricardo} received a Bachelor's and Master's Degree in mechanical and aeronautical engineering (MAE) at the University of California, at Davis in 2008 and 2012 respectively.

He is a doctoral candidate in the department of Mechanical and Aerospace Engineering, University of California at Davis, USA. Before pursuing his doctorate, he worked in commercial aviation as a Structural Analyst, where he led the structural analysis on the Nacelle Inlet for NASAs Quiet Technology Demonstrator 3 (QTD3).
\endbio
\bio{soltani}{Soltani, Iman} received the Ph.D. degree in mechanical engineering from the Massachusetts Institute of Technology (MIT) in 2015. He is a faculty member in the department of Mechanical and Aerospace Engineering, and a graduate faculty in CS, and ECE departments at the University of California, Davis. Previously, he led the Advanced Automation Laboratory at Ford Motor Company. He holds over 17 patents and 40 journal and conference publications in AI, controls, and robotic systems.
\endbio
	
\end{document}